\documentclass[aps,twocolumn]{revtex4-1}

\usepackage{graphicx}
\usepackage[justification=justified,width=\linewidth]{caption}
\usepackage{subcaption}
\usepackage{amsmath}
\usepackage{amsfonts}
\usepackage{amsthm}
\usepackage{amssymb}
\usepackage{amsbsy}
\usepackage{wasysym}
\usepackage{bm}
\usepackage{mathrsfs}
\usepackage{color}
\usepackage{times}
\usepackage[resetlabels]{multibib}

\begin{document}

\title{Two-Stage Ordering Kinetics in Binary Mixtures of Ellipsoidal Particles} % Force line breaks with \\
\author{Parameshwaran A$^{1}$, Sanjay Puri$^{2}$, and Bhaskar Sen Gupta$^{3}$}
\email{bhaskar@iiserbpr.ac.in}

\affiliation{$^{1}$Department of Physics, Vellore Institute of Technology, Vellore, Tamil Nadu 632014, India}

\affiliation{$^{2}$School of Physical Sciences, 
	Jawaharlal Nehru University, New Delhi 110067, India}
	
\affiliation{$^{3}$Department of Physical Sciences, Indian Institute of Science Education and Research Berhampur, Ganjam, Odisha 760003, India}	

\date{\today}% It is always \today, today,
            
\begin{abstract}
We investigate the nonequilibrium phase-ordering kinetics of a binary
mixture of uniaxial ellipsoidal particles interacting via the
anisotropic Gay-Berne potential using large-scale molecular dynamics
simulations. Following a temperature quench from the isotropic and homogeneous phase
into the coexistence region, the system exhibits two distinct ordering
processes occurring on different time scales. At first, rapid orientational ordering leads to the formation and coarsening of nematic domains, characterized by dynamical scaling, generalized Porod-law
behavior, and a growth law $\ell(t) \sim t^{1/2}$ consistent with
nonconserved order-parameter dynamics. Later, the system
undergoes binary phase separation driven by compositional fluctuations.
This compositional ordering process shows a transition from diffusive growth
$\ell(t) \sim t^{1/3}$ to hydrodynamic regimes, with finite-size
scaling analysis indicating asymptotic viscous growth $\ell(t) \sim t$.
Despite the presence of global orientational order, the phase-separation
kinetics remains effectively isotropic. These results demonstrate a
clear separation between orientational and compositional ordering
dynamics, governed by distinct conservation laws, and provide a unified
framework for understanding phase ordering in anisotropic particle
systems.
\end{abstract}

\maketitle

\section{Introduction}

Phase separation and phase ordering kinetics are fundamental nonequilibrium phenomena that arise in a wide variety of physical systems, including binary fluids, polymer blends, alloys, and liquid crystalline materials \cite{Binder1977,Siggia1979,Furukawa1985,SanMiguel1985,Tanaka1995,Beysens2009,Tanaka2011,Kendon2001,Puri1992,Datt2015,Laradji1996,Thakre2008,Ahmad2010, Bray, Puri}. When a homogeneous system is quenched into a thermodynamically unstable or metastable region of its phase diagram, it evolves toward equilibrium via the formation and growth of domains rich in the coexisting phases. The kinetics of this process, commonly referred to as coarsening or domain growth, is governed by factors such as conservation laws, dimensionality, hydrodynamic interactions, and the symmetry of the underlying order parameter.

Although phase separation in systems composed of spherical particles has been extensively studied and is relatively well understood, considerably less attention has been devoted to systems with anisotropic constituents. In such systems, the shape of the particle \cite{Nguyen} introduces additional orientational degrees of freedom, leading to a complex interplay between the positional and orientational ordering. This interplay gives rise to rich phase behavior, including isotropic, nematic, and smectic phases, and significantly affects the pathways through which the system approaches equilibrium. Experimental studies of liquid-crystal mixtures have also demonstrated the rich nonequilibrium kinetics associated with anisotropic constituents. In particular, phase separation in polymer--liquid-crystal mixtures has been observed experimentally, with the evolution of domain morphology and growth kinetics reflecting the interplay between compositional and orientational ordering \cite{Golemme1998,Demyanchuk2006}.

Liquid crystalline systems provide a platform for exploring these effects \cite{Stephen1974,deGennes1995,Singh2000,Priestly2012,Andrienko2018}. In particular, systems of rod-like (calamitic) particles exhibit a sequence of phase transitions from a high-temperature isotropic phase to an intermediate nematic phase with long-range orientational order and, at lower temperatures, to smectic phases characterized by partial translational ordering. The nonequilibrium kinetics associated with these transitions involve the formation and evolution of topological defects, as well as the coupling between orientational alignment and density fluctuations.

Recent simulation studies based on anisotropic interaction models \cite{Billeter,Zannoni,Brown,NBirdi}, such as the Gay-Berne (GB) potential, have provided significant insight into both equilibrium phase behavior and ordering kinetics in liquid crystalline systems. In particular, it has been shown that the anisotropy in the interaction energy plays a crucial role in determining the stability of different phases and the mechanisms of domain growth. For example, domain coarsening in nematic phases is typically governed by curvature-driven dynamics and defect annihilation, leading to a characteristic growth law $L(t) \sim t^{1/2}$ \cite{Allen}, while more complex behavior can arise in phases with additional compositional order.

Despite these advances, the kinetics of phase separation in binary mixtures of anisotropic particles remains relatively unexplored. In such systems, phase separation driven by compositional differences can occur simultaneously with, or subsequent to, orientational ordering. This raises several important questions: How do orientational and compositional ordering processes interact? Do they occur on different time scales? How does particle anisotropy influence the morphology and growth laws of phase-separated domains?

A related issue that has received limited attention is whether the presence of global orientational order influences the universality class of the subsequent phase-separation dynamics or whether the conserved and nonconserved ordering processes remain effectively decoupled. In particular, it is not clear whether nematic alignment can induce anisotropic domain growth or modify the scaling behavior typically observed in binary fluid phase separation. Addressing these questions is essential for developing a unified understanding of phase-ordering kinetics in systems with internal orientational degrees of freedom.

In this work, we address these questions by investigating phase separation and nematic ordering in a binary mixture of ellipsoidal particles that interact via the GB potential. Using large-scale molecular dynamics simulations, we examined the nonequilibrium evolution of the system following a temperature quench from the isotropic phase into the coexistence region. We find that the system exhibits the following ordering process: the development of the nematic order, and the phase separation into compositionally distinct domains. Our results demonstrate that these two processes occur on well-separated time scales and are governed by distinct physical mechanisms, while the late-stage phase separation remains effectively isotropic despite the presence of global orientational order.

To systematically analyze these phenomena, we separately characterize (i) the kinetics of orientational ordering and nematic domain growth and (ii) the compositional ordering associated with binary phase separation. We employ standard tools such as the two-point correlation function, the structure factor, and domain-size scaling analysis to quantify the coarsening dynamics. Our results demonstrate dynamical scaling behavior in the orientational sector and reveal the coexistence of multiple ordering mechanisms operating on different time scales.

The remainder of the paper is organized as follows. In Sect.~II, we describe the model and simulation methodology. In Sect.~III, we present results on the equilibrium phase behavior. Section IV is devoted to the kinetics of orientational ordering and nematic domain growth, followed by an analysis of binary phase separation. Finally, we summarize our findings and discuss their implications in Sec.~V.

\section{Numerical Model and Methods}

\subsection{Interaction Model: GB Potential}

In the present study, we consider a binary system consisting of classical uniaxial ellipsoidal particles of type A and B confined in a three-dimensional simulation box of volume $V$. The particles interact via the anisotropic GB potential, which is a generalized Lennard-Jones (LJ) potential that incorporates both shape anisotropy and orientation-dependent interaction strength \cite{Gay}. This model has been widely used as a prototypical description of liquid crystalline systems, as it successfully captures isotropic, nematic, and smectic phases.
\begin{figure}[h]
	\centering

        \includegraphics[width=0.45\textwidth,height=0.35\textwidth]{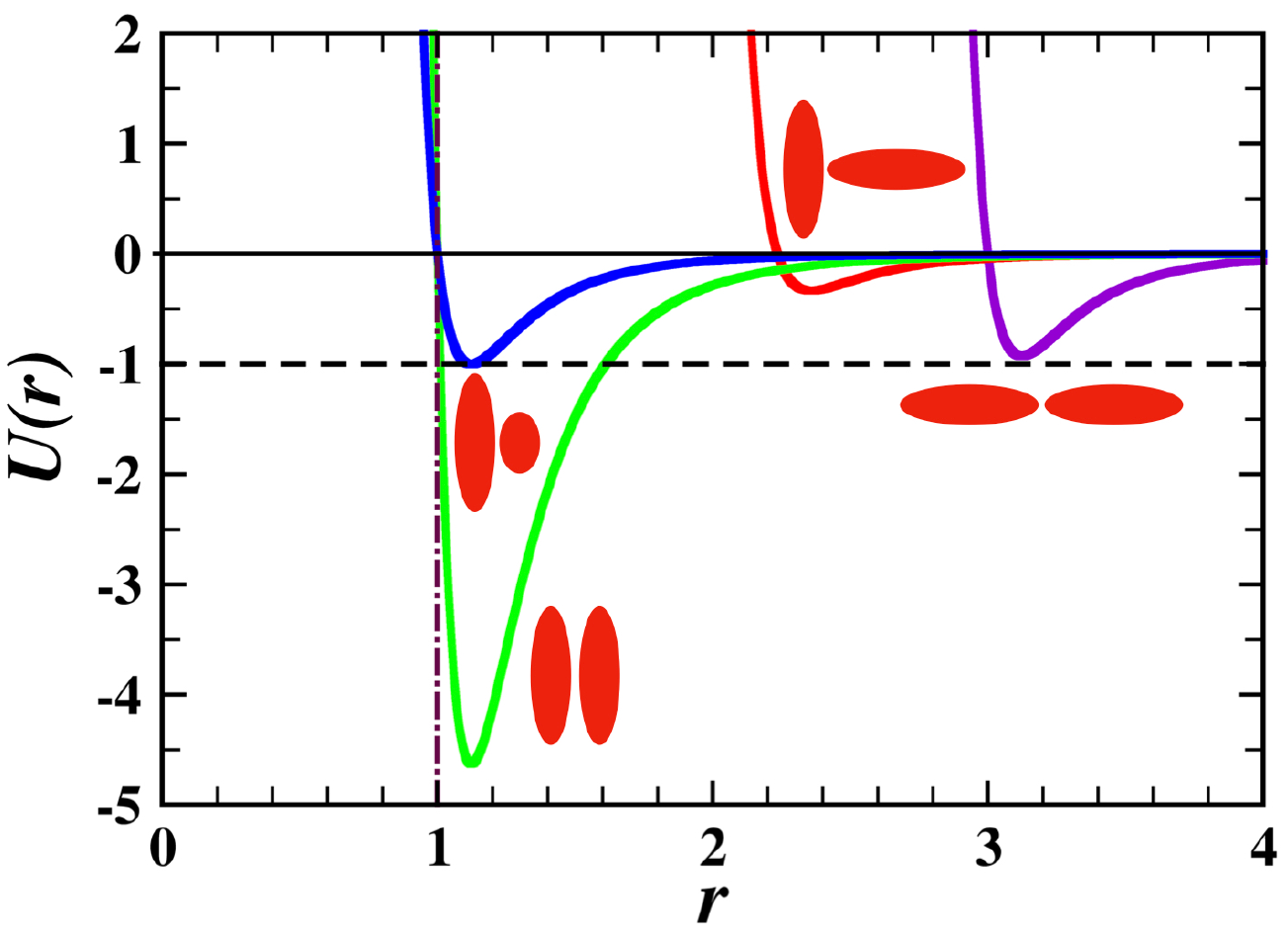}
	\caption{Representative form of the GB interaction potential illustrating the anisotropic dependence of the interaction energy on intermolecular separation and relative orientation. The potential exhibits distinct well depths for side-by-side and end-to-end configurations, reflecting the underlying energy anisotropy of ellipsoidal particles.}
	\label{fig:pot}
\end{figure}

The pair interaction between particles $i$ and $j$ is given by
\begin{equation}
	\begin{aligned}
		U(\mathbf{u}_i,\mathbf{u}_j,\mathbf{r}_{ij}) &=
		4\epsilon(\mathbf{u}_i,\mathbf{u}_j,\hat{\mathbf{r}}_{ij})
		\Bigg\{
		\left[
		\frac{\sigma_0}{r_{ij}-\sigma(\mathbf{u}_i,\mathbf{u}_j,\hat{\mathbf{r}}_{ij})+\sigma_0}
		\right]^{12} \\
		&\quad -
		\left[
		\frac{\sigma_0}{r_{ij}-\sigma(\mathbf{u}_i,\mathbf{u}_j,\hat{\mathbf{r}}_{ij})+\sigma_0}
		\right]^{6}
		\Bigg\},
	\end{aligned}
\end{equation}
where $\sigma_0$ is the characteristic length scale and $\hat{\mathbf{r}}_{ij}$ is the unit vector along the intermolecular separation vector $\mathbf{r}_{ij}$, with $r_{ij}=|\mathbf{r}_i-\mathbf{r}_j|$. The first term represents steep short-range repulsion that prevents particle overlap, while the second term corresponds to the attractive long-range van der Waals interaction.

The quantity $\sigma(\mathbf{u}_i,\mathbf{u}_j,\hat{\mathbf{r}}_{ij})$ denotes the orientation-dependent contact distance between two ellipsoidal particles:
\begin{equation}
	\begin{aligned}
		\sigma (\mathbf{u}_i, \mathbf{u}_j,\hat{\mathbf{r}})
		&=
		\sigma_0
		\left\{
		1
		-
		\frac{\chi}{2}
		\left[
		\frac{(\mathbf{u}_i \cdot \hat{\mathbf{r}} + \mathbf{u}_j \cdot \hat{\mathbf{r}})^2}
		{1 + \chi(\mathbf{u}_i \cdot \mathbf{u}_j)}
		\right.\right. \\
		&\qquad \left.\left.
		+
		\frac{(\mathbf{u}_i \cdot \hat{\mathbf{r}} - \mathbf{u}_j \cdot \hat{\mathbf{r}})^2}
		{1 - \chi(\mathbf{u}_i \cdot \mathbf{u}_j)}
		\right]
		\right\}^{-1/2}.
	\end{aligned}
\end{equation}

Here, the parameter $\chi$ quantifies the anisotropy in the shape of the particles:
\begin{equation}
	\chi=\frac{\kappa^2-1}{\kappa^2+1},
\end{equation}
where $\kappa=\sigma_e/\sigma_s$ is the aspect ratio of the ellipsoids, with $\sigma_e$ and $\sigma_s$ denoting the contact distances end-to-end and side-by-side, respectively. Higher values of $\kappa$ promote orientational ordering as a result of the increase in shape anisotropy.

The orientation-dependent depth of the well $\epsilon(\mathbf{u}_i,\mathbf{u}_j,\hat{\mathbf{r}}_{ij})$ determines the strength of the attractive interaction:
\begin{equation}
	\epsilon(\mathbf{u}_i,\mathbf{u}_j,\hat{\mathbf{r}}_{ij})
	=
	\epsilon_0 \, \epsilon_1^{\mu} \epsilon_2^{\nu},
\end{equation}
where $\epsilon_0$ sets the overall energy scale and the exponents $\mu$ and $\nu$ control the degree of anisotropy in the attractive interactions.

The function $\epsilon_1$ accounts for the dependence of the interaction strength on the relative orientation of the particles with respect to their line of centers:
\begin{equation}
	\epsilon_1
	=
	1
	-
	\frac{\chi'}{2}
	\left\{
	\frac{(\hat{\mathbf{r}}_{ij}\cdot\mathbf{u}_i + \hat{\mathbf{r}}_{ij}\cdot\mathbf{u}_j)^2}
	{1+\chi'(\mathbf{u}_i\cdot\mathbf{u}_j)}
	+
	\frac{(\hat{\mathbf{r}}_{ij}\cdot\mathbf{u}_i - \hat{\mathbf{r}}_{ij}\cdot\mathbf{u}_j)^2}
	{1-\chi'(\mathbf{u}_i\cdot\mathbf{u}_j)}
	\right\},
\end{equation}
while $\epsilon_2$ captures the dependence on the relative alignment of the principal molecular axes:
\begin{equation}
	\epsilon_2 =
	\left[1-\chi^{2}(\mathbf{u}_i\cdot\mathbf{u}_j)^2\right]^{-1/2}.
\end{equation}

The parameter $\chi'$ represents the anisotropy in the attractive interaction:
\begin{equation}
	\chi' = \frac{\kappa'^{1/\mu}-1}{\kappa'^{1/\mu}+1},
\end{equation}
where $\kappa'=\epsilon_s/\epsilon_e$ is the ratio of the well depths for side-by-side and end-to-end configurations. Thus, $\kappa'$ controls the energetic preference for parallel alignment, which plays a crucial role in stabilizing liquid crystalline phases.

In general, the GB model is characterized by four key parameters: $\kappa$ (shape anisotropy), $\kappa'$ (energy anisotropy), and exponents $\mu$ and $\nu$. The interplay between these parameters governs both equilibrium phase behavior and nonequilibrium ordering kinetics.

\subsection{Simulation Details}

The above model is implemented in molecular dynamics (MD) simulations using the LAMMPS package \cite{Plimpton,LAMMPS,Brown}. All quantities are expressed in reduced LJ units. Simulations are performed in a cubic box with periodic boundary conditions applied in all three spatial directions, thereby minimizing finite-size and surface effects.

The system consists of $N = 262144$ uniaxial ellipsoidal particles with a number density $\rho = N/V = 0.3$, corresponding to the dimensions of the box $L_x = L_y = L_z \approx 96$. A binary mixture was prepared by randomly assigning $50\%$ of the particles as type A and the remaining $50\%$ as type B, ensuring an initially homogeneous composition.

Unless otherwise stated, the GB parameters were chosen as $\kappa = 3.0$, $\kappa' = 5.0$, $\mu = 1.0$, and $\nu = 3.0$. These parameters are known to produce a stable nematic phase over a broad temperature range and have been used widely in previous studies \cite{Berardi}. Both species were assigned identical masses ($m=1.0$) and shapes with principal axes $(\sigma_x,\sigma_y,\sigma_z)=(1.0,1.0,3.0)$, corresponding to uniaxial prolate ellipsoids.

The initial particle orientations were generated using random quaternions, ensuring a completely isotropic configuration. Translational and rotational velocities were assigned from a Maxwell-Boltzmann distribution at an initial temperature $T=6.0$. The interaction parameters were chosen as $\epsilon_{0_{AA}}=1.0$, $\epsilon_{0_{BB}}=1.0$, and $\epsilon_{0_{AB}}=0.5$, thus introducing a driving force for phase separation between species. A cutoff radius of $4\sigma_0$ was used for the interaction potential \cite{NBirdi}.  For the present choice of interaction parameters, alignment of the ellipsoidal particles and segregation of the two species are energetically favorable. Since orientational and compositional ordering represent distinct degrees of freedom, four possible states can in general be identified: isotropic-homogeneous (IH), isotropic-segregated (IS), nematic-homogeneous (NH), and nematic-segregated (NS). The equilibrium state depends on the interaction parameters, density, and temperature, and can therefore correspond to any of these states. In this work, we focus on the nonequilibrium ordering dynamics following a quench into the parameter regime in which the NS state is the equilibrium state, and do not consider the kinetics associated with the other possible final states.

The equations of motion for both translational and rotational degrees of freedom are integrated using the velocity Verlet algorithm with a time step $\Delta t = 0.001$. The Nosé-Hoover thermostat (which preserves hydrodynamics) is used to maintain the temperature at the desired value. Neighbor lists were constructed using a cell-based binning method with a skin distance of $1.0$ to improve computational efficiency \cite{Frenkel,Nosé,Verlet,Binder}.

The system is first equilibrated in the canonical ensemble ($NVT$) at $T=6.0$, corresponding to the isotropic phase. This ensures that both compositional and orientational degrees of freedom are fully randomized. Subsequently, the equilibrated configuration is quenched instantaneously to $T=2.3$, which lies within the coexistence region of the phase diagram (see Sect. III A). Such a deep quench drives the system far from equilibrium and triggers both orientational ordering and compositional phase separation. Following the quench, the system evolved in the $NVT$ ensemble for additional $2\times10^7$ time steps. This long simulation time allows the system to access the late-stage coarsening regime, where scaling behavior and growth laws can be reliably extracted. All statistical quantities are averaged over $20$ independent simulation runs, each initialized with different random configurations. This ensemble averaging reduces noise and ensures reliable characterization of the coarsening dynamics and phase behavior.

\section{Results}

For the temperature considered in the present work, we observe that the
system exhibits two distinct types of phase ordering during its
evolution toward equilibrium. At first, the system undergoes a
rapid orientational ordering process, leading to the formation of a
nematic phase characterized by long-range orientational correlations.
 Later, a compositional ordering process sets in, resulting in
binary phase separation between the two species of particles. These two
processes occur on well-separated time scales and are governed by
different physical mechanisms.

Because these two ordering mechanisms involve different conservation
laws and dynamical processes, we organize the discussion of our results
into two separate subsections. In the first subsection, we focus on the
kinetics of orientational ordering and the development of nematic
domains. In the second subsection, we analyze the compositional ordering
associated with binary phase separation and the growth of
compositionally segregated regions.

\subsection{Equilibrium Phase Behavior}

Before discussing the kinetics of ordering, it is useful to first
establish the equilibrium phase behavior of the system. This provides
a reference framework for selecting appropriate quench temperatures
and for interpreting the subsequent nonequilibrium dynamics.

\begin{figure}[h]
	\centering
	\includegraphics[width=0.45\textwidth]{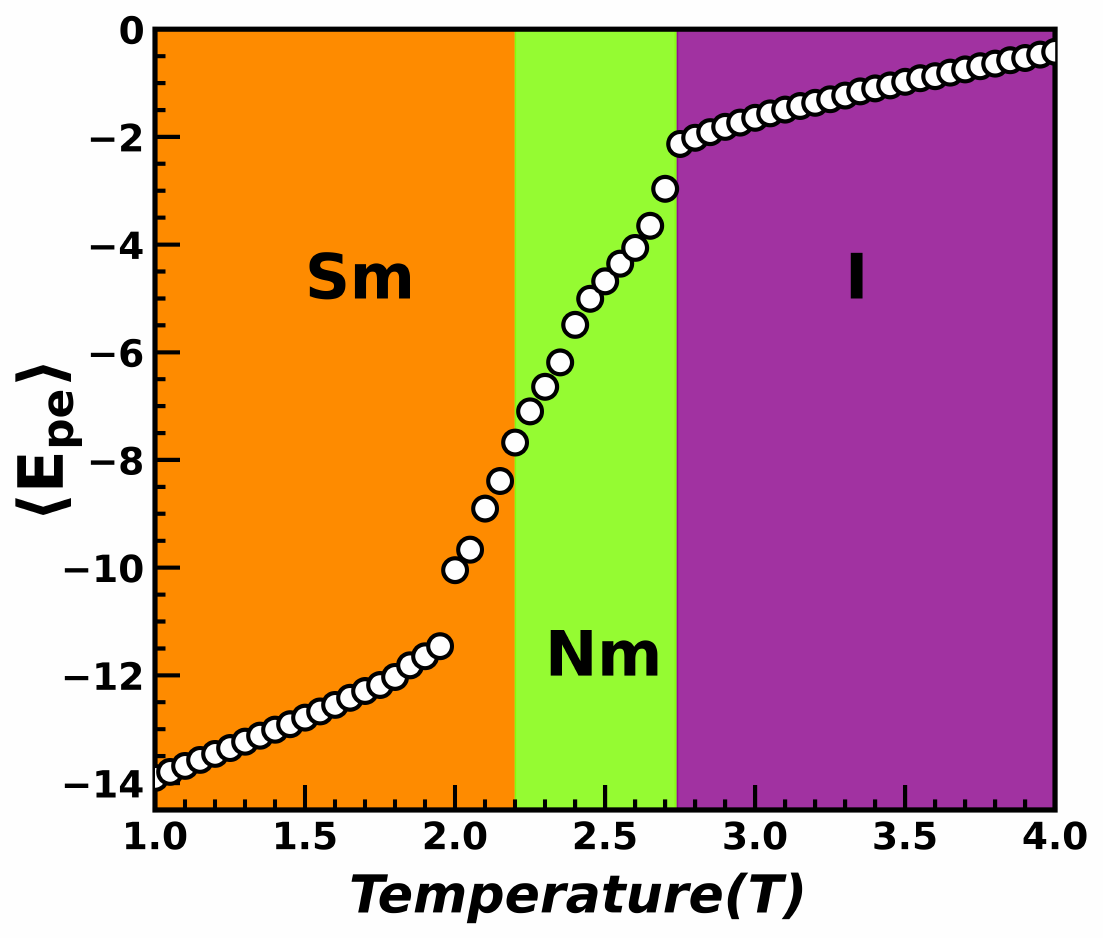}
	\caption{Variation of the average potential energy per particle,
		$\langle E_{pe} \rangle$, as a function of temperature $T$ for the
		GB model system. The discontinuities in the curve mark the
		isotropic-nematic and nematic-smectic phase transitions.}
	\label{fig:phase}
\end{figure}

\begin{figure}[h]
	\centering
	\includegraphics[width=0.45\textwidth]{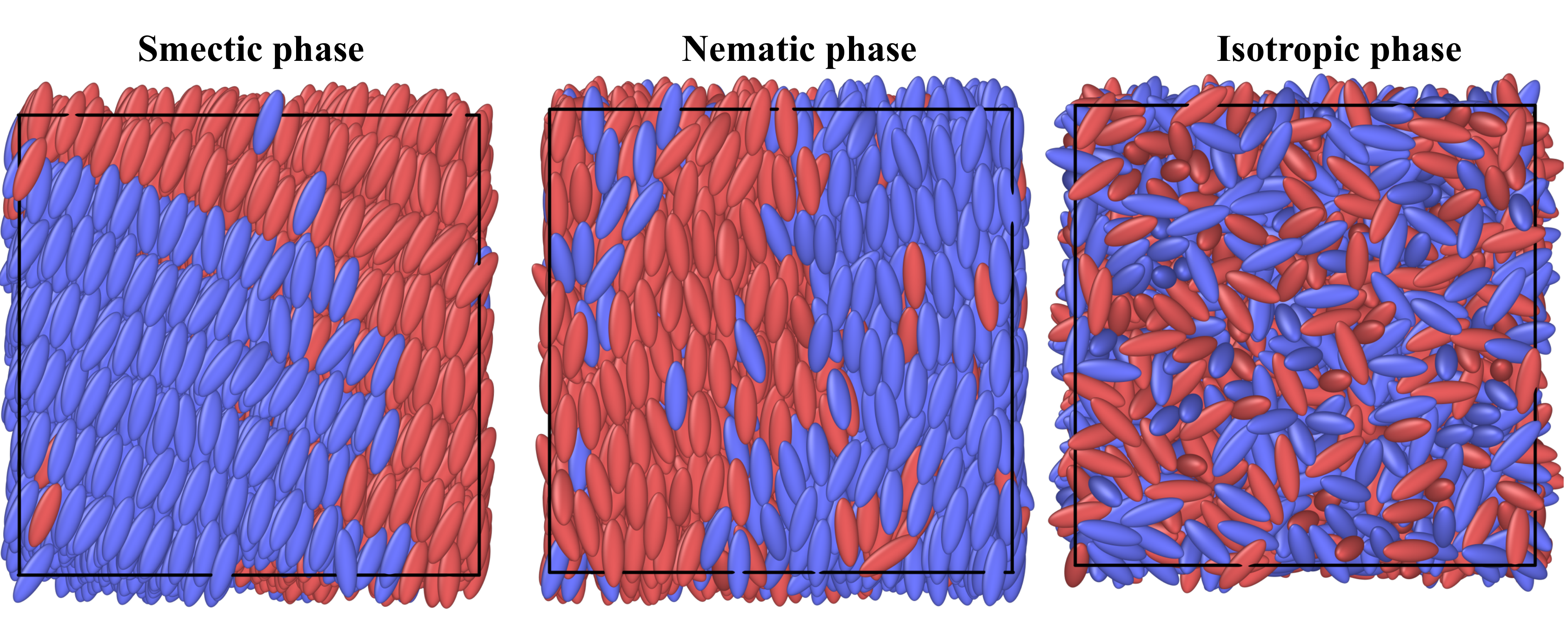}
	\caption{Representative configurations of the GB system in
		different equilibrium phases: isotropic (disordered orientations),
		nematic (long-range orientational order), and smectic (layered
		structure with partial compositional order).}
	\label{fig:different_phase}
\end{figure}

Figure~\ref{fig:phase} shows the variation of the average potential
energy per particle, $\langle E_{pe} \rangle$, as a function of
temperature after the system reaches the equilibrium. The transition temperatures $T_c^{1} \approx2.75$ (corresponding to the isotropic-nematic, $I \rightarrow Nm$, transition) and $T_c^{2} \approx  2.2$
(corresponding to the nematic-smectic, $Nm \rightarrow Sm$, transition)
are identified from the discontinuities observed in
$\langle E_{pe} \rangle$. On the basis of this thermodynamic signature,
three distinct phases can be clearly distinguished: a low-temperature
smectic phase (Sm), an intermediate nematic phase (Nm), and a
high-temperature isotropic phase (I). Representative snapshots of these phases obtained from our simulations are shown in Fig.~\ref{fig:different_phase}.

\begin{figure}[h]
	\centering
	\includegraphics[width=0.45\textwidth]{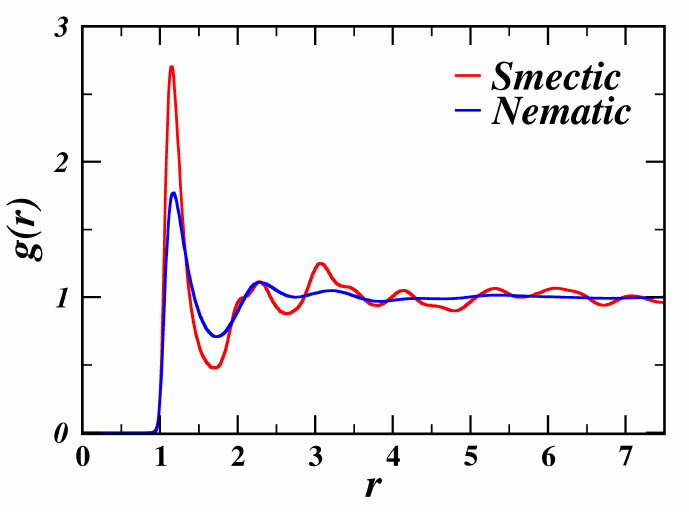}
	\caption{Radial distribution function $g(r)$ for different phases.
		The smectic phase exhibits pronounced oscillations due to layered
		compositional ordering, whereas the nematic phase shows rapidly damped
		oscillations, indicating short-range compositional correlations.}
	\label{fig:pair_cor}
\end{figure}

At low temperatures ($T \lesssim 2.2$), the system exhibits a strongly
ordered smectic phase characterized by both orientational order and
partial compositional layering of the ellipsoidal particles. This phase
corresponds to relatively low potential energy as a result of favorable
particle alignments and efficient packing within layers. As the
temperature increases, the system undergoes a transition to a nematic
phase in the temperature range $2.2 \lesssim T \lesssim 2.75$. In this
regime, the particles retain long-range orientational order, while the
translational order is largely lost. At still higher temperatures
($T \gtrsim 2.75$), thermal fluctuations destroy the orientational
correlations and the system enters the isotropic phase, where both
orientational and compositional orders vanish.

\subsubsection{Characterization of Orientational Order}

To further characterize the structural changes associated with the
isotropic-nematic ($I \rightarrow Nm$) transition, we compute the
nematic order parameter (also known as the $Q$ tensor) \cite{Gennes} using the
orientation vectors $\mathbf{u}_i$ of the ellipsoidal particles.

The orientation vector of each particle is obtained from the quaternion
representation of its rotational state. In the simulation, particle
orientations are stored in terms of quaternions
$\mathbf{q}_i = (q_{0i}, q_{xi}, q_{yi}, q_{zi})$, which describe the
rotation of the particle relative to a reference molecular axis.
Starting from the initial molecular axis $\mathbf{e}_0=(0,0,1)$, the
actual orientation of the particle $i$ is obtained by applying the rotation
operator associated with the quaternion \cite{AllenTildesley2017},
\begin{equation} 
	\mathbf{u}_i = \mathbf{A}(\mathbf{q}_i)\mathbf{e}_0 ,
\end{equation}
where $\mathbf{A}(\mathbf{q}_i)$ is the rotation matrix corresponding
to the quaternion $\mathbf{q}_i$, 
\begin{equation}
\mathbf{A} =
\begin{pmatrix}
1-2q_{yi}^{\,2}-2q_{zi}^{\,2} &
2q_{xi}q_{yi}-2q_{0i}q_{zi} &
2q_{xi}q_{zi}+2q_{0i}q_{yi}
\\[4pt]
2q_{xi}q_{yi}+2q_{0i}q_{zi} &
1-2q_{xi}^{\,2}-2q_{zi}^{\,2} &
2q_{yi}q_{zi}-2q_{0i}q_{xi}
\\[4pt]
2q_{xi}q_{zi}-2q_{0i}q_{yi} &
2q_{yi}q_{zi}+2q_{0i}q_{xi} &
1-2q_{xi}^{\,2}-2q_{yi}^{\,2}
\end{pmatrix}.
\end{equation} 
and $\mathbf{u}_i$ is the unit vector
along the principal molecular axis of the particle $i$. 

Subsequently, these orientation vectors are used to construct the tensor $Q_{\alpha\beta}$, defined as \cite{Gennes}
\begin{equation} 
	Q_{\alpha\beta} = \frac{1}{N}\sum_{i=1}^{N}
	\left(
	\frac{3}{2} u_{i\alpha}u_{i\beta} - \frac{1}{2}\delta_{\alpha\beta}
	\right),
	\label{eq:Qab}
\end{equation}
where $u_{i\alpha}$ and $u_{i\beta}$ are the $\alpha$ and $\beta$
components of $\mathbf{u}_i$, $\delta_{\alpha\beta}$ is the Kronecker
delta, and $N$ is the total number of particles.

The tensor $Q_{\alpha\beta}$ quantifies the degree of orientational
ordering in the system. The largest eigenvalue of $Q_{\alpha\beta}$
defines the global nematic order parameter $S_G$, which measures the
degree of alignment of the particles. In the isotropic phase,
the orientations are randomly distributed and $S_G \approx 0$, whereas
in a perfectly aligned nematic phase $S_G \rightarrow 1$. 

\subsubsection{Characterization of positional Order}

To characterize the structural changes associated with the
nematic-smectic ($Nm \rightarrow Sm$) transition, we compute the
radial distribution function $g(r)$ \cite{Frenkel,Bates}, shown in
Fig.~\ref{fig:pair_cor}. It is defined as
\begin{equation}
	g(r) = \frac{1}{\rho N}
	\left\langle 
	\sum_{i=1}^{N} \sum_{j \ne i}^{N} 
	\delta \left(r - |\mathbf{r}_i - \mathbf{r}_j|\right)
	\right\rangle,
\end{equation}
where $\rho = N/V$ is the number density and $\langle \cdot \rangle$
denotes an ensemble average.

The smectic phase exhibits pronounced oscillations in $g(r)$,
characterized by a strong first peak followed by multiple well-defined
secondary peaks. These features indicate significant compositional
correlations that arise from the layered arrangement of the particles. In
contrast, the nematic phase shows a comparatively weaker first peak
and rapidly damped oscillations \cite{NBirdi}, indicating that compositional
correlations remain short-range despite the presence of long-range
orientational order. This behavior is consistent with the known
structural properties of nematic liquid crystals.

It is important to note that for the coarsening studies considered
here, only approximate phase boundaries are required, since the
quenches are performed sufficiently far from the transition
temperatures. Guided by the equilibrium phase diagram, the temperature
$T=2.3$ was chosen for the kinetic investigations presented in the
following sections, as it lies well within the nematic regime while
remaining sufficiently close to the coexistence region to allow for
subsequent phase separation.

\subsection{Orientational Ordering and Nematic Domain Growth}

\subsubsection{Domain Morphology and Evolution}

Having established the equilibrium phase behavior of the system in the
previous subsection, we now investigate the kinetics of orientational
ordering following a temperature quench in the nematic regime.
Specifically, we focus on the evolution and coarsening of the nematic domains that emerge at early times as the system relaxes toward equilibrium.

Following the quench, the initially random orientations of
the ellipsoidal particles progressively align as a result of anisotropic
interactions encoded in the GB potential. This alignment leads
to the formation of locally ordered nematic domains. As time evolves,
these domains grow and coarsen through the annihilation and
rearrangement of orientational defects, eventually resulting in larger
regions of aligned particles.

\begin{figure}[h]
	\centering
    	\includegraphics[width=0.45\textwidth,height=0.2\textwidth]{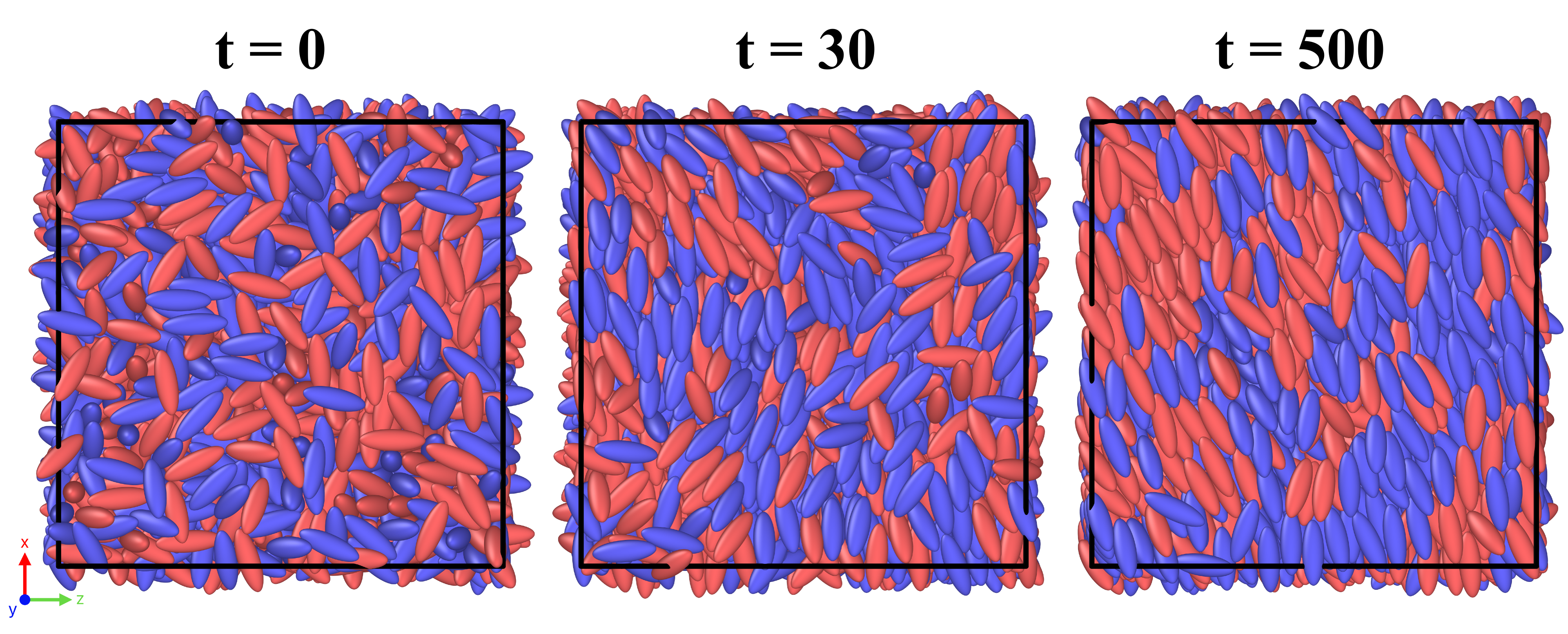}
	\caption{Time evolution of orientational ordering following a
		temperature quench from the isotropic phase ($T=6.0$) to the nematic
		phase ($T=2.3$). The snapshots show the progressive emergence,
		growth, and coalescence of nematic domains. For clarity, only a $20^3$ subvolume taken
		from one corner of the full simulation box is displayed. }
	\label{fig:time_evol}
\end{figure}

Figure~\ref{fig:time_evol} illustrates the temporal evolution of the
nematic ordering process. Immediately after the quench, the system
remains largely disordered. As the simulation progresses, small regions
of locally aligned particles begin to nucleate and grow. Later, these domains merge and coarsen, leading to the development of extended
nematic regions characterized by strong orientational alignment. This
behavior is a hallmark of phase-ordering kinetics in systems with a
nonconserved order parameter, where domain growth is driven by the
reduction of interfacial free energy and the elimination of topological
defects such as disclinations.

\subsubsection{Correlation Function and Dynamical Scaling}

For a translationally invariant system, a standard probe to characterize
the evolving morphology is the spherically averaged two-point equal-time
correlation function \cite{Parameshwaran,Sen Gupta,Davis}
\begin{equation}\label{eq:cr}
	C(r,t)= \langle S_L(0,t)S_L(\vec{r},t) \rangle
	- \langle S_L(0,t)\rangle \langle S_L(r,t)\rangle ,
\end{equation}
where $S_L$ is the coarse-grained orientational order
parameter. The order parameter field is constructed using a
coarse-graining procedure in which the simulation box is divided into
non-overlapping sub-boxes of size $(3.0\sigma_0)^3$. This choice ensures
that each cell contains approximately $8$-$10$ particles, thereby
providing a meaningful local average. In this manner, the continuum
liquid-crystal configuration is mapped onto a simple cubic lattice
consisting of $(32)^3$ cells \cite{NBirdi}.

The relevant local order parameter used to characterize orientational
ordering is the second-rank Legendre polynomial, commonly referred to as
the local nematic order parameter \cite{Callan-Jones},
\begin{equation}
	S_L = \left\langle \frac{3\cos^2\theta_i - 1}{2} \right\rangle .
\end{equation}
Here, $\theta_i$ is the angle between the principal axis of an
ellipsoidal particle located in the $i$th sub-box and the global
nematic director $\mathbf{n}$. The director $\mathbf{n}$ is obtained
from the eigenvector corresponding to the largest eigenvalue of the
nematic order tensor $Q_{\alpha\beta}$ defined in
Eq.~\ref{eq:Qab}. Physically, $\mathbf{n}$ represents the preferred
orientation of the particles and serves as a reference axis for
quantifying alignment.

The angular brackets denote an average over all particles contained
within a given sub-box. Thus, $S_L$ provides a measure of the degree of
local orientational ordering. For a completely isotropic configuration,
$S_L \approx 0$, while $S_L \rightarrow 1$ for perfectly aligned
particles. Intermediate values correspond to partial ordering,
characteristic of the nematic phase. This coarse-grained field is used
both for visualization of nematic domains and for quantitative analysis
of their growth.

\begin{figure}[h]
	\centering
	\includegraphics[width=0.45\textwidth,height=0.4\textwidth]{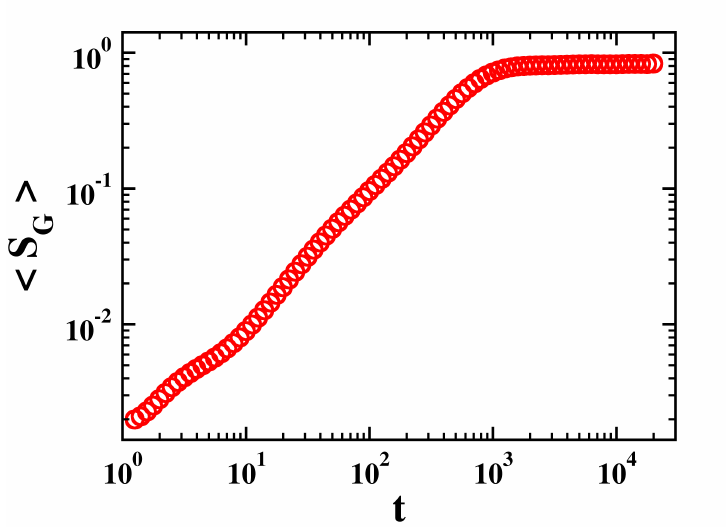}
	\caption{Time evolution of the global nematic order parameter $S_G$}
	\label{fig:SG_vs_t}
\end{figure}
To further quantify the kinetics of orientational ordering, we examine
the temporal evolution of the global nematic order parameter $S_G$. 
Figure~\ref{fig:SG_vs_t} shows the variation of
$S_G$ as a function of the time following the quench. Starting from a
disordered isotropic configuration with $S_G \approx 0$, the system
initially exhibits a rapid increase in $S_G$, indicating the
rapid establishment of long-range orientational correlations. The
order parameter subsequently saturates to a steady value
characteristic of the nematic phase. This rapid saturation clearly
demonstrates that orientational ordering occurs on a much shorter
time scale compared to the subsequent compositional ordering (discussed in the next section) associated with binary phase separation, thereby providing direct evidence for the separation of time scales in the ordering dynamics.

For systems undergoing phase-ordering kinetics, evolving structures
often exhibit dynamical scaling at late times. In this regime, the
equal-time correlation function satisfies \cite{ Bray}
\begin{equation}
	C(r,t) \equiv \tilde{C}\left(\frac{r}{\ell(t)}\right),
\end{equation}
where $\tilde{C}$ is a time-independent scaling function and $\ell(t)$
is the characteristic domain size. This scaling form implies that the
morphology at different times is statistically self-similar when lengths
are rescaled by $\ell(t)$, so that the entire time dependence is encoded
in the growth of this length scale.

In the present study, the domain size $\ell(t)$ is extracted from the
decay of $C(r,t)$ by defining it as the distance at which the
correlation function first crosses the value $0.2$. This operational
definition provides a robust and widely used estimate of the average
domain size during coarsening.

\begin{figure}[h]
	\centering
    	\includegraphics[width=0.45\textwidth,height=0.4\textwidth]{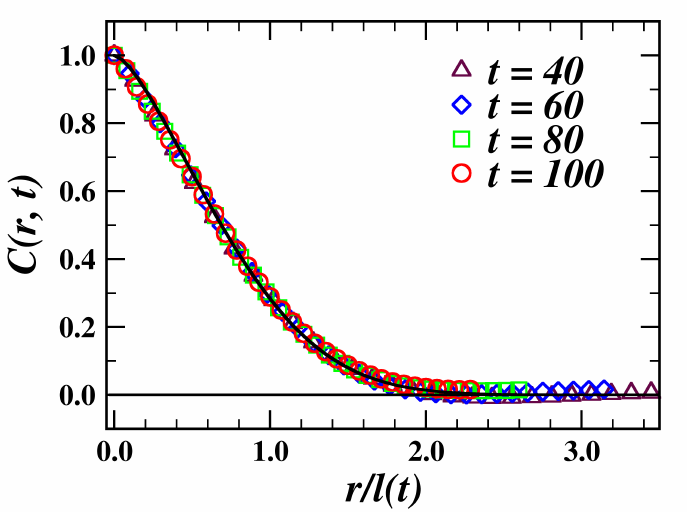}
	\caption{Scaled orientational correlation function $C(r,t)$ plotted as a
		function of the scaled distance $r/\ell(t)$ at different times
		following the quench. The collapse of the data onto a single master
		curve demonstrates dynamical scaling and statistical self-similarity
		during nematic domain growth. The solid line represents the
		Bray-Puri-Toyoki (BPT) theoretical prediction for a nonconserved
		order parameter with $n=2$ components.}
	\label{fig:orient_cor}
\end{figure}

Figure~\ref{fig:orient_cor} shows the scaled correlation function $C(r,t)$ plotted as a function of $r/\ell(t)$ at different times. The data collapse into a single master curve, providing strong evidence for dynamical scaling in the orientational ordering process~\cite{NBirdi}.
This result indicates that, although the characteristic domain size
increases with time, the overall morphology remains statistically
invariant when expressed in terms of the scaled variable $r/\ell(t)$.  

To further quantify this scaling behavior, we compare our numerical
results with the Bray-Puri-Toyoki (BPT) function~\cite{Puri}, which
provides an analytical form for the correlation function in systems with a nonconserved $n$-component order parameter. The solid line in Fig.~\ref{fig:orient_cor} corresponds to this theoretical prediction, given by
\begin{equation}
	\begin{aligned}
		C(r,t) &= \langle S_L(0,t)S_L(r,t) \rangle \\
		&= \frac{n\gamma}{2\pi} \left[ B\left(\frac{n+1}{2}, \frac{1}{2}\right) \right]^2 
		F\left(\frac{1}{2}, \frac{1}{2}; \frac{n+2}{2}; \gamma^2 \right),
	\end{aligned}
	\label{eq:BPT}
\end{equation}
where $B(x,y)$ denotes the Beta function and $F(a,b;c;z)$ is the
hypergeometric function. The scaling variable $\gamma$ is defined as
$\gamma = e^{-x^2}$, with $x = r/\ell(t)$. For the present system,
$n=2$, corresponding to string (vortex-line) defects in three
dimensions. We find that the analytical form shows excellent agreement
with the simulation data, thereby providing further support for the
dynamical scaling scenario. 

\subsubsection{Structure Factor and Generalized Porod Law}

Further insight into the evolving morphology can be obtained from the
structure factor, defined as the Fourier transform of $C(r,t)$ \cite{Bray,Bhattacharyya1,Bhattacharyya2},
\begin{equation}\label{eq:sk}
	S(k,t) = \langle |\psi(\vec{k},t)|^2 \rangle
	= \int C(\vec{r},t) e^{i\vec{k}\cdot\vec{r}} d\vec{r},
\end{equation}
where $k$ is the wave vector and $\psi(\vec{k},t)$ is the Fourier
transform of the order-parameter field. In the large-$k$ limit, the
structure factor exhibits a power-law decay consistent with the
generalized Porod law \cite{Puri},
\begin{equation}
	S(k,t) \sim k^{-(d+n)},
\end{equation}
where $d$ is the spatial dimensionality and $n$ is the number of
components of the order parameter. This behavior arises from scattering
off interfaces and topological defects in the orientational field.

The structure factor provides complementary information to the
real-space correlation function by highlighting characteristic length
scales in reciprocal space. In particular, the peak of $S(k,t)$ shifts
to smaller $k$ with increasing time, reflecting the growth of domains.

For self-similar evolution, the structure factor obeys the scaling form
\begin{equation}
	S(k,t) \equiv \ell(t)^d \tilde{S}(k\ell(t)),
\end{equation}
where $\tilde{S}$ is a time-independent scaling function. Thus, plotting
$S(k,t)/\ell(t)^d$ against $k\ell(t)$ should yield a collapse of data
from different times.

\begin{figure}[h]
	\centering
	    	\includegraphics[width=0.45\textwidth,height=0.4\textwidth]{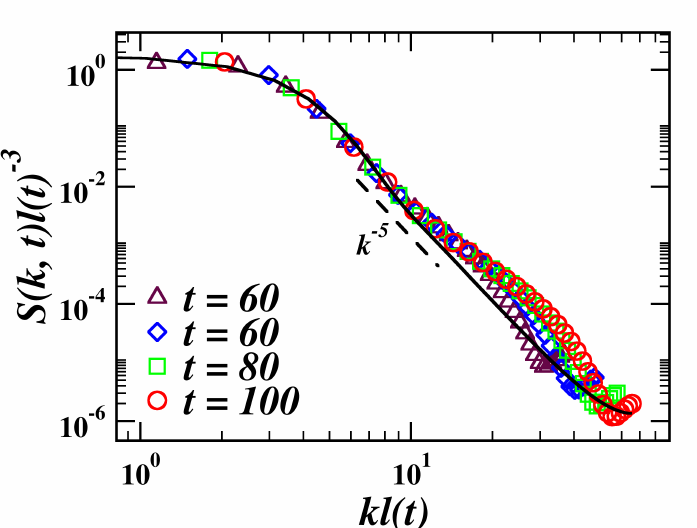}
	\caption{Scaled structure factor $S(k,t)/\ell(t)^d$ plotted as a function of
		the scaled wave number $k\ell(t)$ at different times following the
		quench. The collapse of the data onto a single master curve confirms
		dynamical scaling during nematic domain growth. The dashed line
		indicates the generalized Porod-law decay, $S(k,t) \sim k^{-5}$, in
		the large-$k$ regime. The solid line corresponds to the Fourier
		transform of the Bray-Puri-Toyoki correlation function [Eq.~\ref{eq:BPT}].}
	\label{fig:orient_struc}
\end{figure}

Figure~\ref{fig:orient_struc} shows the scaled structure factor at
different times. The excellent collapse further confirms the validity of
the dynamical scaling hypothesis. In the asymptotic large-$k$ regime,
the structure factor follows $S(k,t) \sim k^{-5}$ for $d=3$ and $n=2$,
consistent with scattering from string-like topological defects in the
nematic field \cite{Blundell,Banerjee}.

For further validation, we compare the scaled structure factor with
the Fourier transform of the Bray-Puri-Toyoki (BPT) correlation
function [Eq.~\ref{eq:BPT}]. As shown in Fig.~\ref{fig:orient_struc},
the theoretical prediction is in good agreement with the simulation
data over a wide range of wave numbers, including the crossover to
the Porod regime. This agreement is consistent with the expected
$n=2$ vector order parameter for nematic ordering in three dimensions.
\subsubsection{Growth Law}

\begin{figure}[h]
	\centering
    	\includegraphics[width=0.45\textwidth,height=0.5\textwidth]{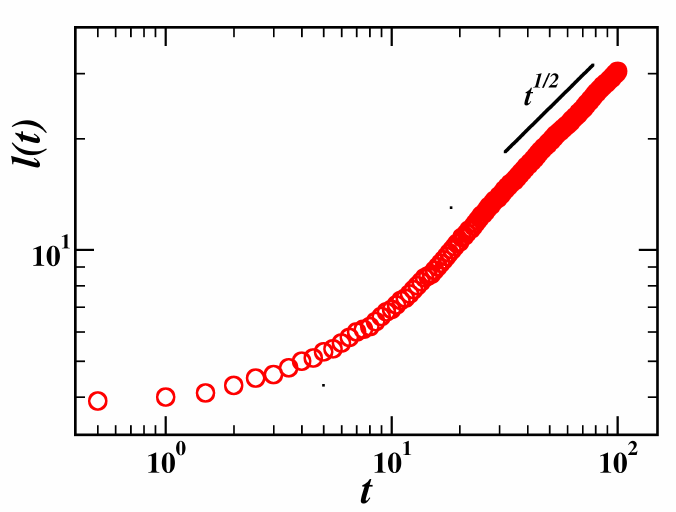}
	\caption{Log-log plot of the characteristic domain size $L(t)$ versus
		time $t$. The solid line indicates the growth law $L(t) \sim t^{1/2}$,
		consistent with curvature-driven coarsening in systems with a
		nonconserved order parameter.}
	\label{fig:orient_lengthscale}
\end{figure}

Finally, we examine the growth law for the characteristic domain size.
Figure~\ref{fig:orient_lengthscale} shows $\ell(t)$ as a function of time on
a log-log scale. After an initial transient regime associated with the
formation of local order, the system enters a scaling regime where
\[
\ell(t) \sim t^{1/2}.
\]
This behavior is consistent with the Lifshitz-Allen-Cahn mechanism \cite{Allen}, which governs curvature-driven coarsening in systems with a
nonconserved order parameter. In this regime, domain growth is driven by
the annihilation of topological defects. 

The system sizes and simulation times used in the present study are large enough to clearly establish the $t^{1/2}$ growth law for nematic ordering in the GB model.

\subsection{Compositional Ordering and Binary Phase Separation}

\subsubsection{Morphology and Evolution}

While the preceding subsection focused on the kinetics of orientational
ordering leading to the formation and coarsening of nematic domains,
the system also exhibits a slower structural evolution associated with
compositional ordering. In particular, the binary mixture of ellipsoidal
particles undergoes phase separation driven by the difference in
interaction strengths between like and unlike species. As the system
evolves within the nematic regime, particles of the same type tend to
aggregate, giving rise to the formation of compositionally distinct
domains. This process corresponds to binary phase separation, where the
system gradually evolves toward a state characterized by spatially
segregated regions rich in either component. 

Unlike the relatively fast development of orientational order, the
compositional ordering process occurs on longer time scales and is
governed by the diffusion and rearrangement of particles. The emerging
patterns display domain structures whose characteristic length scale
increases with time as the system approaches equilibrium. This clear
separation of time scales allows one to distinguish between the rapid
establishment of orientational order and the comparatively slow
coarsening of compositional domains.

\begin{figure}[h]
	\centering
	    	\includegraphics[width=0.45\textwidth,height=0.2\textwidth]{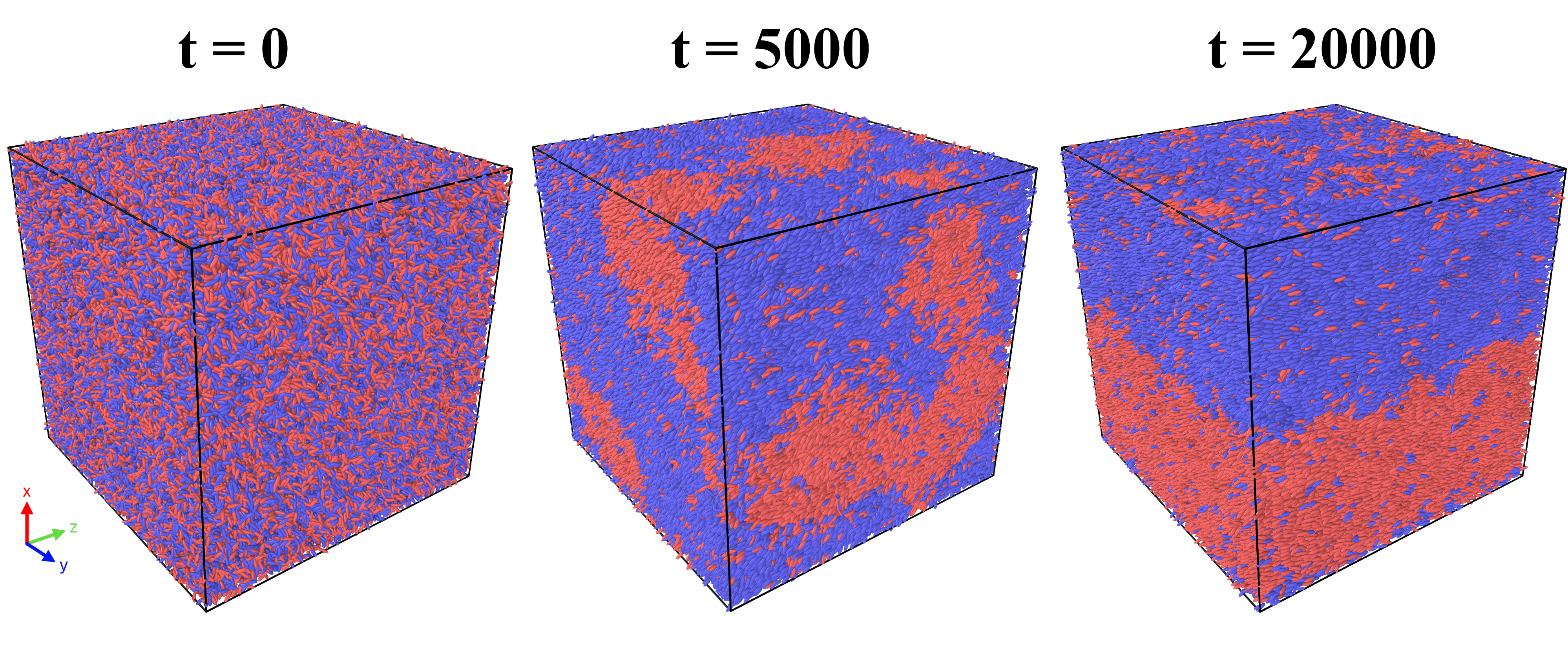}
	\caption{Time evolution of the composition field following a quench
		from the isotropic phase ($T=6.0$) to the nematic regime
		($T=2.3\epsilon/k_B$). The snapshots show the progressive emergence,
		growth, and coarsening of domains enriched in the two particle species.}
	\label{fig:time_evol_2}
\end{figure}

Figure~\ref{fig:time_evol_2} shows a sequence of representative
snapshots that illustrate the temporal evolution of domain structures
after the quench. Immediately after the quench, the system is
spatially homogeneous with no significant compositional ordering.
As time progresses, fluctuations in the local composition are
amplified, leading to the emergence of small domains enriched in
either component. These domains subsequently coarsen through diffusion
and rearrangement of particles, resulting in the formation of larger
compositionally segregated regions as the system evolves toward
equilibrium. 

\subsubsection{Correlation Function and Structure Factor}

To quantitatively characterize the compositional ordering and
binary phase separation process, we analyze the spatial
distribution of particle species using the two-point
correlation function $C(r,t)$ and the corresponding
structure factor $S(k,t)$ associated with the composition
field following Eqs.~\ref{eq:cr} and~\ref{eq:sk}, respectively. These quantities provide complementary information
on the evolving morphology and the characteristic
length scale of segregated domains.

The order parameter $\psi(\vec{r},t)$ is defined from
the coarse-grained local density difference between the
two species of particles within a cubic box of size $(3\sigma)^3$
centered at position $\vec{r}=(x,y,z)$. Specifically,
$\psi(\vec{r},t)$ is assigned the value $+1$ if species $A$
dominate within the box and $-1$ if species $B$ dominate \cite{Bhattacharyya2}.
This binary representation allows us to map the microscopic
particle configuration onto a lattice-like scalar order
parameter field that clearly distinguishes the two phases.
Using this field, the equal-time spatial correlation
function $C(r,t)$ is computed.

\begin{figure}[h]
	\centering
    	\includegraphics[width=0.45\textwidth,height=0.4\textwidth]{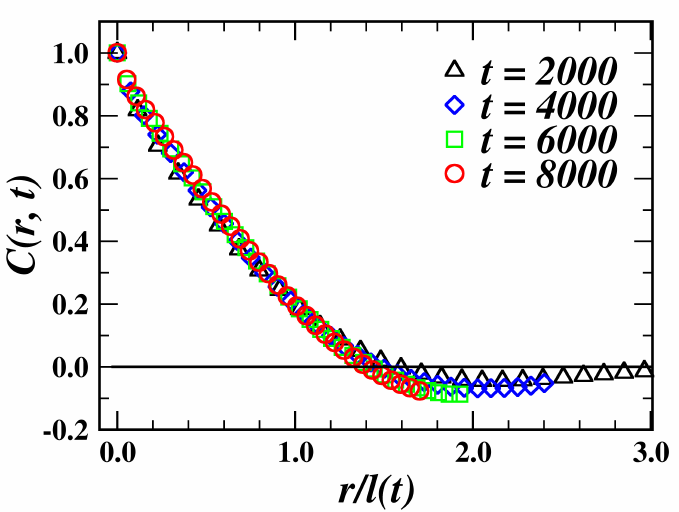}
	\caption{Scaled two-point correlation function $C(r,t)$ of the
		composition field plotted as a function of $r/\ell(t)$ at different
		times. The data collapse onto a single master curve demonstrates
		dynamical scaling during the phase-separation process.}
	\label{fig:binscal_cor}
\end{figure}

Figure~\ref{fig:binscal_cor} shows the scaled correlation function
$C(r,t)$ plotted against $r/\ell(t)$ at different times.
The excellent collapse of the data into a single master curve
indicates that the domain morphology obeys dynamical scaling
during the coarsening process. This implies that the evolving
structures are statistically self-similar and can be characterized
by a single time-dependent length scale $\ell(t)$.

Further insight into domain morphology is obtained from the structure factor $S(k,t)$,
\begin{figure}[h]
	\centering
	   	\includegraphics[width=0.45\textwidth,height=0.4\textwidth]{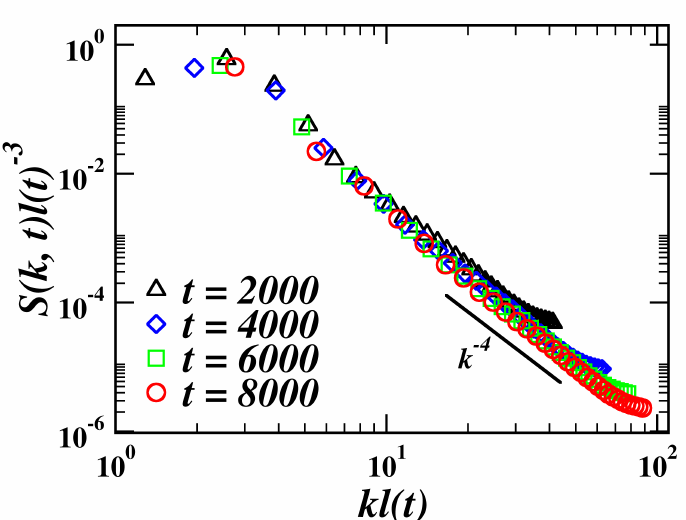}
	\caption{Scaled structure factor $S(k,t)/\ell(t)^d$ plotted as a
		function of $k\ell(t)$ at different times. The collapse of the data
		confirms dynamical scaling, while the large-$k$ decay is consistent
		with Porod-law behavior arising from sharp interfaces between domains.}
	\label{fig:binscal_struc}
\end{figure}
Figure~\ref{fig:binscal_struc} shows the corresponding scaled
structure factor $S(k,t)\ell(t)^{-d}$ plotted as a
function of $k\ell(t)$ at different times. The collapse
of the curves again confirms the validity of the
dynamical scaling hypothesis for the binary phase
separation process. In the large-$k$ regime, the structure factor follows
the Porod-law behavior $S(k,t) \sim k^{-(d+1)}$, which in three dimensions corresponds to $S(k,t) \sim k^{-4}$. This decay is due to the scattering
from sharp interfaces between domains rich in species $A$ and $B$, indicating the presence of well-defined interfaces during the late stages of
phase separation. 

\subsubsection{Growth Laws and Hydrodynamic Effects}

\begin{figure}[h]
	\centering
	
        	\includegraphics[width=0.55\textwidth,height=0.45\textwidth]{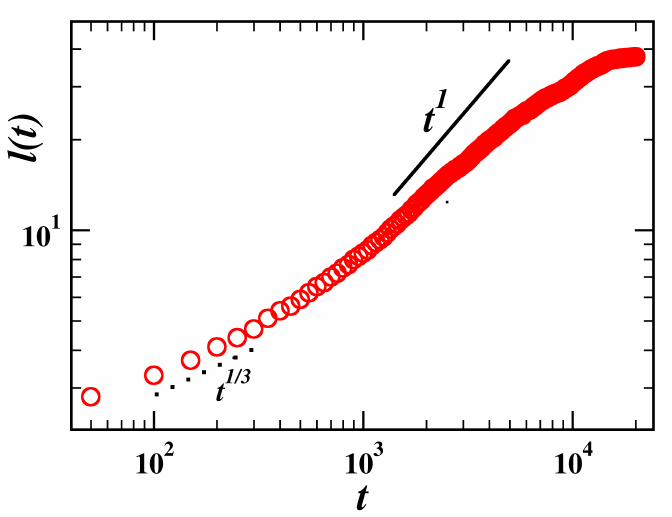}
	\caption{Log-log plot of the characteristic domain size $\ell(t)$
		as a function of time. }
	\label{fig:bin_lengthscale}
\end{figure}

Finally, we examine the temporal evolution of the characteristic
domain size $\ell(t)$ associated with the binary phase separation
process. The length scale $\ell(t)$ is extracted from the first
zero-crossing of the correlation function $C(r,t)$. The resulting
variation of $\ell(t)$ with time is shown in Fig.~\ref{fig:bin_lengthscale}
on a log-log scale.

At first, the system exhibits a growth regime consistent with
the classical diffusive coarsening mechanism \cite{Lifshitz,Wagner}, characterized by the Lifshitz-Slyozov law $\ell(t) \sim t^{1/3}$, which is typical for phase separation in systems with a conserved order parameter.
In this regime, domain growth is primarily governed by diffusive
transport of particles across interfaces that separate the two phases.

As time progresses, the growth rate increases and the system
enters a regime where the characteristic domain
size follows approximately $\ell(t) \sim t^{1}$ (with a deviation). This behavior
suggests the onset of hydrodynamic effects that accelerate the
coarsening process through advective transport of particles.
However, at very late times a deviation from this power law
is observed.

\begin{figure}[h]
	\centering
        	\includegraphics[width=0.45\textwidth,height=0.4\textwidth]{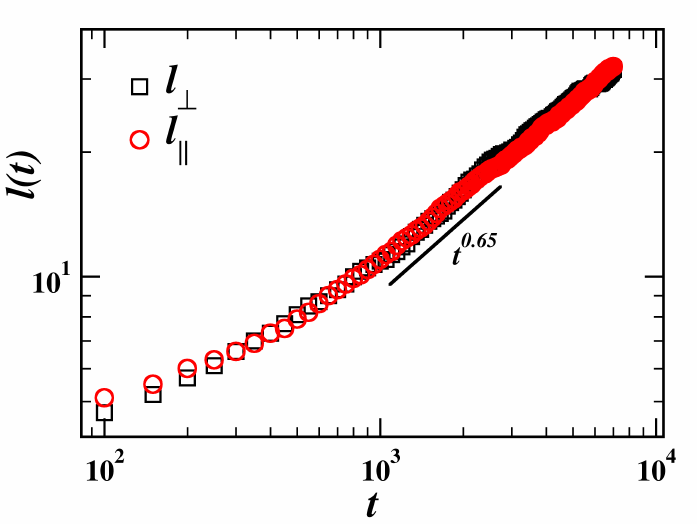}
	\caption{Time evolution of the characteristic domain size measured
		parallel and perpendicular to the global nematic director
		$\mathbf{n}$. The nearly identical growth behavior along both
		directions indicates that phase separation remains effectively
		isotropic despite the presence of nematic order.}
	\label{fig:parallel_vs_perpendicular}
\end{figure}

To investigate whether this deviation originates from anisotropic
domain growth induced by the underlying nematic ordering, we
examined the evolution of domain sizes separately along directions
parallel and perpendicular to the global nematic director
$\mathbf{n}$. Figure~\ref{fig:parallel_vs_perpendicular}
shows the time evolution of the domain size measured parallel
to the nematic director and perpendicular to it.

As seen in Fig.~\ref{fig:parallel_vs_perpendicular}, the domain
growth along the two directions exhibits nearly identical
power-law behavior throughout the simulation time window. This
observation indicates that the compositional phase separation
process remains effectively isotropic and is not significantly
affected by the preferred orientational direction of the
nematic phase. Therefore, the change in the effective
growth exponent at very late times cannot be attributed to
anisotropic domain growth.

Instead, the observed deviation from the $t^{1}$ law
is most likely a consequence of finite-size effects that arise
when the growing domains become comparable to the size of the system.
In this regime, domain coarsening is constrained by the finite
simulation box, which naturally leads to a slowdown in growth dynamics. Such finite-size-induced deviations from the asymptotic scaling law are commonly observed in large-scale simulations of phase-separating systems. Therefore, the observed exponent in
Fig.~\ref{fig:bin_lengthscale} should be interpreted as
pre-asymptotic, finite-size values arising from crossover effects
between early-time diffusion-dominated growth and the late-time
viscous hydrodynamic regime, rather than as universal exponent.
\subsubsection{Finite-Size Effects and Scaling Analysis}
To obtain a reliable estimate of the growth exponent, we
performed a finite-size scaling (FSS) analysis. 
\begin{figure}[h]
	\centering
        	\includegraphics[width=0.45\textwidth,height=0.4\textwidth]{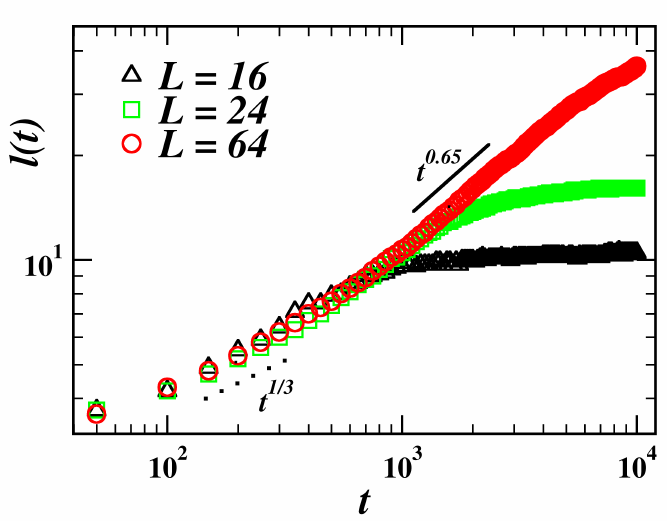}
	\caption{Time evolution of the characteristic domain size $\ell(t)$
		for different system sizes $L = 16$, $24$, and $64$. Larger systems
		exhibit an extended scaling regime before the onset of finite-size
		saturation, enabling a more accurate determination of the growth
		exponent.}
	\label{fig:bin_lengthscale_diff_L}
\end{figure}
Figure~\ref{fig:bin_lengthscale_diff_L} shows the temporal
evolution of the characteristic length scale $\ell(t)$ for
three different system sizes, $L = 16$, $24$, and $64$.
It is evident that the largest system ($L=64$) displays the
most extended power-law regime before the onset of saturation,
highlighting the importance of studying sufficiently large
system sizes to reliably capture the asymptotic domain growth
behavior. In contrast, smaller systems exhibit an earlier
deviation from scaling due to finite-size constraints. However,
simulations of very large systems are computationally demanding,
making it necessary to employ systematic analysis techniques to
accurately determine the growth exponent from finite-size data.

Within the FSS framework \cite{Bray,bubble,Binder}, the growth of domains after a quench can be described by separating an initial transient regime from the asymptotic scaling regime. Let $t_0$ denote the crossover
time beyond which the system exhibits self-similar coarsening.
The characteristic length scale can then be expressed as
\begin{equation}
	\ell(t) = \ell_0 + A (t - t_0)^{\alpha},
\end{equation}
where $\ell_0 = \ell(t_0)$ represents the characteristic
length scale at the beginning of the scaling, $A$ is a nonuniversal
amplitude and $\alpha$ is the domain growth exponent.

To systematically analyze finite-size effects, we introduce a
dimensionless scaling function $Y(X)$ defined as
\begin{equation}
	\ell(t) - \ell_0 = Y(X)\,(\ell_{\max} - \ell_0),
\end{equation}
where $\ell_{\max}$ denotes the maximum domain size attainable for a finite system. The corresponding scaling variable $X$ is given by
\begin{equation}
	X = \frac{(\ell_{\max} - \ell_0)^{1/\alpha}}{t - t_0}.
\end{equation}
In this representation, $Y(X)$ is expected to be independent of
the size of the system. Therefore, data for different system sizes should
collapse onto a single master curve when plotted in terms of
these scaled variables, provided that the correct values of
$\ell_0$, $t_0$, and $\alpha$ are used. Furthermore, in the
asymptotic limit of large $X$, the scaling function is expected
to follow the power-law behavior
\begin{equation}
	Y(X) \sim X^{-\alpha},
\end{equation}
which provides a direct route to estimating the growth exponent.

\begin{figure}[h]
	\centering
        	\includegraphics[width=0.45\textwidth,height=0.4\textwidth]{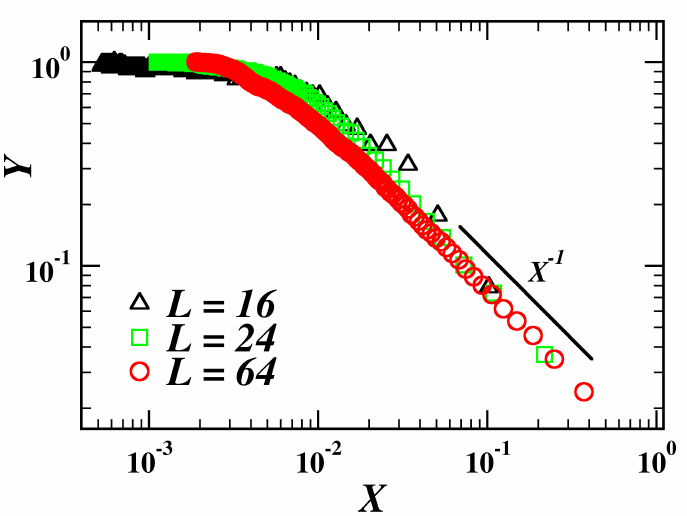}
	\caption{Finite-size scaling collapse of the domain growth data.
		The dimensionless scaling function $Y(X)$ is plotted as a function
		of the scaling variable $X$ for different system sizes. The data
		collapse confirms the validity of the scaling ansatz and allows
		for an accurate estimation of the growth exponent.}
	\label{fig:FSS}
\end{figure}

Figure~\ref{fig:FSS} shows the scaling function $Y(X)$ plotted as a
function of the scaling variable $X$ on a log-log scale. The excellent
collapse of the data obtained for different system sizes demonstrates
the validity of the FSS analysis and confirms the
existence of a well-defined scaling regime in the late stages of domain
coarsening. The best collapse is obtained for the parameters
$\ell_0 = 6.5$, $t_0 = 100$, and the growth exponent $\alpha = 1.0$.

In the large-$X$ regime, the scaling function exhibits a clear
power-law decay consistent with $Y(X) \sim X^{-\alpha}$, with the
exponent $\alpha = 1.0$. This result indicates that the domain growth
in the late-time regime follows a linear growth law 
$\ell(t) \sim t$, which is characteristic of the viscous
hydrodynamic regime. In this regime, domain coarsening is driven
primarily by hydrodynamic flow, where the motion of the interfaces is
controlled by viscous dissipation rather than purely diffusive
transport. The observed exponent is therefore consistent with
theoretical predictions for phase separation dynamics dominated
by hydrodynamic interactions.

\section{CONCLUSIONS}

In this work, we have investigated the nonequilibrium ordering kinetics of a binary mixture of uniaxial ellipsoidal particles interacting through the anisotropic Gay-Berne potential. Following a quench from the isotropic-homogeneous (IH) state into a regime where the nematic-segregated (NS) state is favored, we find that the system approaches the final ordered state through two well-separated ordering processes. Orientational ordering occurs rapidly, leading to the formation and coarsening of nematic domains, while compositional ordering sets in at substantially later times and produces large-scale phase separation. Thus, although the two order parameters are coupled through the underlying particle interactions, their kinetics can remain strongly separated in time.

The orientational ordering exhibits the characteristics of nonconserved order-parameter dynamics. The nematic domains obey dynamical scaling, with the scaled correlation function collapsing onto a common master curve and showing good agreement with the Bray-Puri-Toyoki form. The corresponding structure factor displays the generalized Porod behavior $S(k,t)\sim k^{-5}$ expected for an $n=2$ orientational order parameter in three dimensions. Consistent with these observations, the characteristic nematic domain size grows as $\ell(t)\sim t^{1/2}$, indicating curvature-driven coarsening associated with the annihilation and rearrangement of orientational defects.

The subsequent compositional ordering is governed by the conserved composition field and exhibits a different sequence of growth regimes. At early times, the characteristic domain size follows the diffusive growth law $\ell(t)\sim t^{1/3}$, followed by a crossover to faster growth as hydrodynamic effects become important. Finite-size scaling analysis gives a late-time growth exponent $\alpha\simeq 1$, consistent with viscous hydrodynamic coarsening, $\ell(t)\sim t$. The observed sequence of growth laws therefore reflects the distinct dynamical constraints associated with the two ordering processes: nonconserved orientational relaxation at early times and conserved, increasingly hydrodynamic, compositional evolution at later times.

An important result is that the established nematic order does not produce appreciable anisotropy in the subsequent phase-separation dynamics. The characteristic domain sizes measured parallel and perpendicular to the nematic director exhibit nearly identical temporal behavior. Thus, for the parameters studied here, the orientational order provides the structural background for compositional ordering without changing the large-scale isotropic character of the phase-separation kinetics. The apparent decoupling should therefore be understood as a dynamical decoupling at large length scales rather than as a complete microscopic independence of the two order parameters.

An interesting direction for future work is to investigate systematically how particle anisotropy controls the relative time scales of orientational and compositional ordering. Although such a systematic study is beyond the scope of the present work because of the substantial computational cost, our additional observations (not shown here) indicate that increasing the aspect ratio of the ellipsoidal particles leads to faster orientational ordering while simultaneously slowing down compositional phase separation. Particle anisotropy therefore appears to enhance the separation between the characteristic time scales of the two ordering processes and may provide a means of tuning the kinetic pathway toward the final ordered state. For the parameters considered here, however, the two processes remain largely decoupled at the level of large-scale coarsening, as evidenced by the absence of appreciable anisotropy in the phase-separation dynamics despite the presence of established nematic order. A systematic exploration over a wider range of aspect ratios and interaction parameters would be valuable for determining how particle anisotropy controls the relative kinetics and possible ordering pathways.

Overall, our results establish a clear connection between nematic ordering and conventional phase-separation kinetics in an anisotropic particle system. The emergence of distinct time scales and growth laws demonstrates that multiple ordering processes can proceed sequentially without necessarily altering the large-scale coarsening behavior of one another. These findings provide a useful framework for understanding phase ordering in systems where compositional and orientational degrees of freedom evolve simultaneously but on different dynamical time scales.

\subsection*{Acknowledgements} 
Parameshwaran A. acknowledges DST-SERB, India for doctoral fellowship. S. Puri is grateful to
ANRF, India for financial support via a J.C. Bose fellowship. B. Sen Gupta acknowledges Science and Engineering Research Board (SERB), Department of Science and Technology (DST), Government of India (no. CRG/2022/009343) for financial support.

\end{document}